\documentclass[10pt]{article}

\title{Modeling quantum scattering of fundamental particles by classical, deterministic processes}
\author{Marijan Ribari\v c and Luka \v Su\v ster\v si\v c\\Jo\v zef Stefan Institute, Jamova 39, 1000 Ljubljana, Slovenia\\ e-mail: luka.sustersic@ijs.si}

\begin{document}

\maketitle

\begin{abstract}
We point out that results obtained by M. Ribari\v c and L. \v Su\v ster\v si\v c, hep-th/0403084, and by M. Blasone, P. Jizba and H. Kleinert, quant-ph/0409021, suggest that the path-integral formalism is the key to a derivation of quantum physics from classical, deterministic physics in the four-dimensional space-time. These results and the 't Hooft conjecture, hep-th/0104219, suggest to consider a relativistic, non-material medium, an ether, as a base for non-local hidden-variable models of the physical universe.
\end{abstract}
\vspace{1cm}
\noindent PASC numbers: 03.65.Ta; 03.65.Ud; 04.20.Cv; 12.60.-i; 31.15.Kb

\noindent UDC: 530.145, 530.12, 515.3, 530.145


\vspace{1cm}

We can represent contemporary knowledge about the physical universe by about twenty so-called \it fundamental particles \rm (most of which are taken as experimentally verified). Excepting the graviton, interactions between these particles are described by the standard model in terms of about another forty elementary particles, cf. e.g. \cite{wiki-pp, exper}. The idea that we can model all matter in terms of elementary particles that are not made up of anything smaller dates back at least twenty-six centuries. 

\section{Two weighty problems with the standard model}

According to Dirac \cite{Dirac}, ``One can distinguish between two main procedures for a theoretical physicist. One of them is to work from the experimental basis \ldots The other procedure is to work from the mathematical basis. One examines and criticizes the exiting theory. One tries to pin-point the faults in it and then tries to remove them. The difficulty here is to remove the faults without destroying the very great successes of the existing theory''.

In modeling the quantum scattering of fundamental particles (QSFP), there are two more than sixty years old problems concerning \it conceptual consistency \rm and \it physical completeness \rm of such models: 
\begin{itemize}
\item[(A)]{}When we calculate the observable results of QSFP by a quantum field theory, the perturbative expansions of $n$-point Green functions result in ultravioletly divergent integrals. Formal regularization methods that make these integrals finite are not conceptually consistent with the established concepts of theoretical physics, see 
e.g. \cite{wiki-pp, Dirac} and \cite{Weinberg} subsection 1.3. 
\item[(B)]{}If we go beyond a strictly operational interpretation of QSFP, certain results suggest the existence of arbitrary fast effects. This was brought up already by the Einstein, Podolski and Rosen thought experiment, and was experimentally verified by violations of the Bell inequalities. As far as we know, there is no such model of QSFP and of the associated arbitrary fast effects that is consistent with Einstein's relativity postulates.
\end{itemize}
Inspired by the above Dirac remark, we believe that solving these two theoretical problems should be relevant to our understanding of QSFP. Namely, according to 't Hooft \cite{Hooft}, ``History tells us that if we hit upon some obstacle, even if it looks like a pure formality or just a technical complication, it should be carefully scrutinized. Nature might be telling us something, and we should find out what it is''. 

\section{Should the free fields satisfy wave equations?}

Commenting on the fact that contemporary theories about QSFP grew out of application of the quantization procedure to classical fields that satisfy wave equations, Bjorken and Drell \cite{Bjork} pointed out the following facts about such a procedure which are still as relevant today as forty years ago: ``The first is that we are led to a theory with differential wave propagation. The field functions are continuous functions of continuous parameters $x$ and $t$, and the changes in the fields at a point $x$ are determined by properties of the fields infinitesimally close to the point $x$.

For most wave fields (for example, sound waves and the vibrations of strings and membranes) such a description is an idealization which is valid for distances larger than the characteristic length which measures the granularity of the medium. For smaller distances these theories are modified in a profound way.

The electromagnetic field is a notable exception. Indeed, until the special theory of relativity obviated the necessity of a mechanistic interpretation, physicists made great efforts to discover evidence for such a mechanical description of the radiation field. After the requirement of an ``ether'' which propagates light waves had been abandoned, there was considerably less difficulty in accepting this same idea when the observed wave properties of the electron suggested the introduction of a new field. Indeed there is no evidence of an ether which underlies the electron wave. However, it is a gross and profound extrapolation of present experimental knowledge to assume that a wave description successful at ``large'' distances (that is, atomic lengths $\approx 10^{-8}$ cm) may be extended to distances an indefinite number of orders of magnitude smaller (for example, to less than nuclear lengths $\approx 10^{-13}$ cm).

In the relativistic theory, we have seen that the assumption that the field description is correct in arbitrarily small space-time intervals has led---in perturbation theory---to divergent expressions for the electron self-energy and the bare chage. Renormalization theory has sidestepped these divergence difficulties, which may be indicative of the failure of the perturbation expansion. However, it is widely felt that the divergences are symptomatic of a chronic disorder in the small-distance behaviour of the theory.

We might then ask why local field theories, that is, theories of fields which can be described by differential laws of wave propagation, have been so extensively used and accepted. There are several reasons, including the important one that with their aid a significant region of agreement with observations has been found \ldots But the foremost reason is brutally simple: there exists no convincing form of a theory which avoids diferential field equations''. So what is the way out of this quandary? cf.\cite{mi002}.

\section{Feynman's X-ons}

To our knowledge it was Feynman \cite{Feynman} who first suggested that the basic partial-differential equations of theoretical physics might be actually describing classical, macroscopic motion of some infinitesimal entities he called X-ons. However, such a description by partial-differential equations of fluid dynamics can be improved to take some account of the underlying classical, microscopic motion through the linearized Boltzmann integro-differential transport equations for one-particle distributions \cite{Liboff}. So we may hope that we can improve on the macroscopic description of motion of X-ons by appropriate linear integro-differential equations for functions of the space-time variable, and of an additional four-vector variable. And then use the corresponding Lagrangians as the free-field Lagrangians in theories about QSFP. Thus we propose the following 

\noindent{\bf Premise: } In modelling QSFP by path-integral methods, we will use such relativistic free-field Lagrangians that their Euler-Lagrange equations may be taken as the integro-differential equations for one-particle distribution functions describing, in the linear approximation, the transport of some infinitesimal entities, X-ons.

In contrast, quantum field theories use free-field Lagrangians whose Euler-Lagrange equations (the Klein-Gordon, Maxwell, and Dirac equations) describe differential wave propagation of continuous fields of just the space-time variable.

Starting from this premise, we put forward a finite, relativistic theory of QSFP without using auxiliary particles \cite{mi001}. It shows that in order (A)~to avoid ultraviolet divergencies when computing a perturbative model of QSFP, and (B)~to model faster-than-light effects without introducing the possibility of faster-than-light communication, it suffices to appropriately change only the free-field Lagrangians of the standard model, while retaining their locality in space-time and Lorentz invariance. Using functions of two independent four-vector variables, we based this finite theory on the path-integral formalism in the four-dimensional space-time, and the Lehmann-Symanzik-Zimmermann reduction formula. 

\section{A classical, deterministic theory that implies a finite theory of QSFP}

The mathematical formalism of quantum mechanics provides statistical predictions about QSFP. Which prompted an ongoing philosophical discussion whether we can predict the future, or whether the behaviour of nature is unpredictable and random. According to Einstein, ``God does not play dice with the Universe''. And those agreeing with him are looking for a classical, deterministic theory that would imply quantum-mechanical predictions as a statistical approximation, cf. Genovese \cite{Genovese} for a review of research on hidden-variable theories. In particular, 't Hooft \cite{Hooft2} conjectured that ``We should not forget that quantum mechanics does not really describe what kind of dynamical phenomena are actually going on, but rather gives us probabilistic results. To me, it seems extremely plausible that any reasonable theory for the dynamics at the Planck scale would lead to processes that are so complicated to describe, that one should expect apparently stochastic fluctuations in any approximation theory describing the effects of all of this at much larger scales. It seems quite reasonable first to try a classical, deterministic theory for the Planck domain. One might speculate then that what we call quantum mechanics today, may be nothing else than an ingenious technique to handle this dynamics statistically.''

In their paper \cite{Blasone}, Blasone, Jizba and Kleinert ``have attempted to substantiate the recent proposal of G. 't Hooft in which quantum theory is viewed as not a complete field theory, but is in fact an emergent phenomenon arising from a deeper level of dynamics. The underlying dynamics are taken to be classical mechanics with singular Lagrangians supplied with an appropriate information loss condition. With plausible assumptions about the actual nature of the constraint dynamics, quantum theory is shown to emerge when the classical Dirac-Bergmann algorithm for constrained dynamics is applied to the classical path integral \ldots''. These results, and those obtained by Ribari\v c and \v Su\v ster\v si\v c \cite{mi001} suggest the following

\noindent{\bf Assumption} We will deepen our understanding of the physical universe by modeling its phenomena by using a four-dimensional relativistic medium governed by deterministic laws that imply an appropriate, path-integral based statistical description of QSFP.

The preceding premise suggests we assume that the medium for building non-local hidden-variable theories consists of an infinite number of infinitesimal spin-0, spin-$1/2$, and spin-1 particles, X-ons, with arbitrary four-momenta moving in the three-dimensional space according to deterministic, Poincare-invariant equations of motion, and exerting forces on one another in such a way as to imply a model of QSFP through a path-integral. Let us call it F-medium, with F standing for Feynman who suggested X-ons as the unifying concept for description of physical universe, though he did not specify their properties, cf. the following note by Polyakov.

According to the philosophical point of view of Lorentz, Einstein, Michelson, Dirac and few other theorists \cite{WikiLe}, there should be some non-material space-filling medium, an ether. If possible, the F-medium used for modeling QSFP by classical, deterministic processes should have the propeties expected from such a philosophy-inspired medium.

We might expect the models of the physical universe based on the F-medium to be analogous to the models of kinetic theory, cf.\cite{Liboff}. But, as noted by Polyakov \cite{Polyakov}, ``Elementary particles existing in nature resemble very much excitations of some complicated medium (ether). We do not know the detailed structure of the ether but we have learned a lot about effective Lagrangians for its low energy excitations. It is as if we knew nothing about the molecular structure of some liquid but did know the Navier-Stokes equation and could thus predict many exciting things. Clearly, there are lots of different possibilities at the molecular level leading to the same low energy picture.''

\end{document}